\documentclass[onecolumn,showpacs,preprintnumbers]{revtex4}
\usepackage{dcolumn}
\usepackage{bm}
\usepackage{amsmath}
\usepackage{graphicx}
\usepackage{mathrsfs}
\usepackage{amssymb}
\usepackage{amsfonts}

\begin{document}
\draft
\title{Amplitude and phase control of gain without inversion in a four-level atomic
system using loop-transition}
\author{Jinhua Zou,\footnote{%
Author for correspondence. jhzou@yangtzeu.edu.cn} Dahai Xu and Huafeng Zhang}
\address{Department of Physics, Yangtze University, Jingzhou 434023,\\
People's Republic of China}

\begin{abstract}
Amplitude and phase control of gain without inversion is investigated in a
four level loop-structure atomic system. Two features are presented. One is
that gain without inversion can be obtained through the amplitude control of
the applied fields. The other is that gain without inversion show a
phase-dependence on the relative phase between the fields applied on the two
two-photon transitions. Gain and phase-dependence originate from
interference induced by such a loop-transition structure.
\end{abstract}

\pacs{PACS numbers: 42.50.Gy, 42.50.Lc, 03.67.Mn, 03.65.Ud}
\maketitle

\narrowtext

Gain without inversion corresponds to the gain at the emission peak. For the
case of emission, there is no population inversion between the upper atomic
state and the lower atomic state. If this involved transition is a lasing
transition, then laser without inversion can be obtained $\left[ 1-6\right] $%
. That is to say, gain without inversion may lead to laser without inversion
in such systems. As laser without inversion is an important coherence
phenomena, it has been paid considerable interest $[7-12]$. For a
traditional laser action, population inversion between the laser transition
is needed. Later it was suggested that lasing without inversion can be
realized through interference between different channels $\left[ 7-9\right] $%
. In these schemes, the lasing transition is adjusted by the interference
terms induced by the driving fields or the initial coherence. Through the
variation of the interference, lasing without inversion can be achieved. The
origin of laser without inversion is the gain at the emission peak we
explained before. Gain without inversion has been realized in many system,
such as atomic systems $\left[ 1-6\right] $ and semiconductor nanostructures
$[13]$. Among these schemes, most of the schemes put the effort to obtain
the condition of laser without inversion $[7-12]$ and seldom schemes are
trying to deal with the phase control of the laser without inversion $[6]$.
As the amplitude and phase control is a strong way to adjust the response of
atomic-field system, it may also can be used to control gain without
inversion in certain interacting atom-field systems.

It is acceptable that in a loop structure, the relative phase of the applied
fields does contribute a lot to the response of the probe field. It means
that in such cases, coherent population trapping $[14]$, the absorption
spectra $\left[ 15,16\right] $, the steady state population $\left[
17\right] $, or the spontaneous emission spectra $\left[ 18-19\right] $ will
have a phase-dependence, which can not be found in non-loop transition
structures. There are two ways to form the loop-structure. One is to drive
the dipole-allowed transitions in simple atomic system $[20-21]$. The second
one is to use microwave fields to drive the dipole-prohibited transition
together with the dipole-allowed transitions $[6,17,22]$. Among the three
methods, the first one is the best one if just simple atomic system is
considered. Here we choose the first way to use four level atom with double
middle states, one ground state and one excited state to form a
loop-transition.

In this paper we are going to investigate the amplitude and phase control of
gain without inversion in a four-level loop-structure system. The motivation
lies in the fact that in a loop-structure atomic system, relative phase
plays an essential role in the response. The aim of this paper is to present
the phase control action of the gain without inversion. The considered
loop-structure contains two two-photon transitions, which will be shown
later. The main results are as follows: (i) Gain without inversion can be
obtained by varing the amplitude of applied fields. (ii) Gain without
inversion displays a phase dependence on the relative phase between the
fields applied on the two two-photon transitions. As the phase changes from
0 to $\pi $, gain and absorption zones exchange. Phase-dependent gain
without inversion in such a system originates from interference induced by
two two-photon loop-transition structure. Gain and phase-dependence are
attributed to coherence and interference induced by such a loop-transition.

The considered atomic system with a loop-transition is shown in Fig. 1. Four
coherent driving fields are applied on dipole-allowed transitions
respectively. The transitions can be divided into two kinds of two-photon
transitions: $|1\rangle \leftrightarrow |2\rangle \leftrightarrow |3\rangle $
and $|1\rangle \leftrightarrow |2^{^{\prime }}\rangle \leftrightarrow
|3\rangle $. There are two possible coupling schemes according to the
probing transition, which are shown in Fig. 1(a) and Fig. 1(b). We first
concentrate on the low-level probing case shown in Fig. 1(a). Coherent
fields with frequencies $\omega _{c}$ and $\omega _{p}$ are applied on the
transition $|1\rangle \leftrightarrow |2\rangle \leftrightarrow |3\rangle $,
and coherent fields with frequencies $\omega _{1}$ and $\omega _{2}$ are
applied on the transition $|1\rangle \leftrightarrow |2^{^{\prime }}\rangle
\leftrightarrow |3\rangle $, respectively. We call this case low-level
probing as the probe field is applied on the transition including the lower
level. The Hamiltonian of the whole atom-field system can be written as
\begin{equation}
H=H_{0}+H_{I}
\end{equation}
\begin{equation}
H_{0}=\Delta _{p}\sigma _{22}+(\Delta _{1}+\Delta _{2})\sigma _{33}+\Delta
_{2}\sigma _{2^{^{\prime }}2^{^{\prime }}}
\end{equation}

\begin{equation}
H_{I}=-\Omega _{p}\sigma _{21}-\Omega _{c}\sigma _{32}-\Omega _{1}\sigma
_{32^{^{\prime }}}-\Omega _{2}\sigma _{2^{^{\prime }}1}+h.c.
\end{equation}
where h.c. symbols the hermit conjugate of the front terms. The detunings $%
\Delta _{j}(j=p,c,1,2)$ are defined as the frequency differences between the
applied fields and the corresponding transitions. For specific, $\Delta
_{p}=\omega _{21}-\omega _{p}$, $\Delta _{c}=\omega _{32}-\omega _{c}$, $%
\Delta _{1}=\omega _{32^{^{\prime }}}-\omega _{1}$, and $\Delta _{c}=\omega
_{2^{^{\prime }}1}-\omega _{2}$. In order to fulfill the rotating
transformation the four detunings should satisfy $\Delta _{p}+\Delta
_{c}=\Delta _{1}+\Delta _{2}$, i.e., the two two-photon transitions have the
same sum detuning. When Operators $\sigma _{ij}$ are population operators
when $i=j$ ($j=1,2,2^{^{\prime }},3$), and $\sigma _{ij}$ are flip operators
when $i\neq j$ ($i$, $j=1,2,2^{^{\prime }},3$). And $\Omega _{j}$ ($%
j=p,c,1,2 $) are Rabi frequencies of the coherent driving fields and
generally they are assumed to be complex values. The dynamic behavior of the
system can be described by the master equation of the density matrix $\rho $
as $\left[ 23\right] $%
\begin{equation}
\dot{\rho}=-i[H,\rho ]+L\rho
\end{equation}
where the term $L\rho $ describes the contribution of the atomic decay
terms, and it has the form
\begin{eqnarray}
L\rho &=&\sum \frac{\gamma _{jk}}{2}(2\sigma _{jk}\rho \sigma _{kj}-\sigma
_{kj}\sigma _{jk}\rho -\rho \sigma _{kj}\sigma _{jk}),\text{\quad }
\nonumber \\
jk &=&2^{^{\prime }}3,12^{^{\prime }},23,12
\end{eqnarray}
\begin{figure}[tbph]
\centering
\includegraphics[width=9 cm]{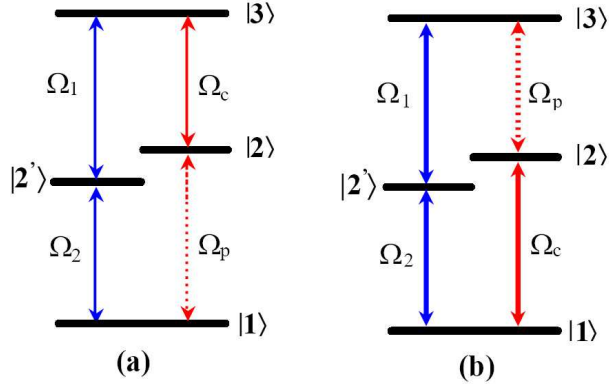}
\caption{(a) Low-level probing scheme for two two-photon transitions
atom-field system. (b) Up-level probing scheme for two two-photon
transitions atom-field system.}
\label{fig1}
\end{figure}
The elements of the density matrix can be obtained directly from the master
equation. It is known that for loop-transition, the relative of the fields
plays important role in the result. The complex Rabi frequencies are defined
as $\Omega _{c}=\Omega _{c}^{0}e^{i\phi _{c}},\Omega _{p}=\Omega
_{p}^{0}e^{i\phi _{p}},\Omega _{1}=\Omega _{1}^{0}e^{i\phi _{1}},\Omega
_{2}=\Omega _{2}^{0}e^{i\phi _{2}}$, where $\Omega _{j}^{0}$($j=p,c,1,2$)
are strength of the Rabi frequencies.. In order to see the relative phase,
we make the transformation as $\rho _{32}=\sigma _{32}e^{i\phi _{c}},\rho
_{32^{^{\prime }}}=\sigma _{32^{^{\prime }}}e^{i\phi _{1}},\rho _{21}=\sigma
_{21}e^{i\phi _{p}},\rho _{2^{^{\prime }}1}=\sigma _{2^{^{\prime
}}1}e^{i\phi _{2}},\rho _{31}=\sigma _{31}e^{i(\phi _{1}+\phi _{2})},\rho
_{22^{^{\prime }}}=\sigma _{22^{^{\prime }}}e^{i(\phi _{1}-\phi _{c})}$ and $%
\rho _{jj}=\sigma _{jj}$ $(j=1,2,2^{^{\prime }},3)$. After the
transformation the motion of the elements are as following

\begin{eqnarray*}
\dot{\sigma}_{33} &=&-(\gamma _{1}+\gamma _{c})\sigma _{33}-i\Omega
_{c}^{0}\sigma _{32}-i\Omega _{1}^{0}\rho _{32^{^{\prime }}}+i\Omega
_{c}^{0}\sigma _{23} \\
&&+i\Omega _{1}^{0}\sigma _{2^{^{\prime }}3} \\
\dot{\sigma}_{22} &=&\gamma _{c}\sigma _{33}-\gamma _{p}\sigma _{22}+i\Omega
_{c}^{0}\sigma _{32}-i\Omega _{p}^{0}\sigma _{21}-i\Omega _{c}^{0}\sigma
_{23} \\
&&+i\Omega _{p}^{0}\sigma _{12} \\
\dot{\sigma}_{2^{^{\prime }}2^{^{\prime }}} &=&\gamma _{1}\sigma
_{33}-\gamma _{2}\sigma _{2^{^{\prime }}2^{^{\prime }}}+i\Omega
_{1}^{0}\sigma _{32^{^{\prime }}}-i\Omega _{2}^{0}\sigma _{2^{^{\prime
}}1}-i\Omega _{1}^{0}\sigma _{2^{\prime }3} \\
&&+i\Omega _{2}^{0}\sigma _{12^{^{\prime }}} \\
\dot{\sigma}_{32} &=&-\Gamma _{32}\sigma _{32}-i\Omega _{c}^{0}(\sigma
_{33}-\sigma _{22})-i\Omega _{p}^{0}e^{-i\phi }\sigma _{31} \\
&&+i\Omega _{1}^{0}\sigma _{2^{^{\prime }}2} \\
\dot{\sigma}_{32^{^{\prime }}} &=&-\Gamma _{32^{^{\prime }}}\sigma
_{32^{^{\prime }}}-i\Omega _{1}^{0}(\sigma _{33}-\sigma _{2^{^{\prime
}}2^{^{\prime }}})-i\Omega _{2}^{0}\sigma _{31}+i\Omega _{c}^{0}\sigma
_{22^{^{\prime }}} \\
&&+i\Omega _{p}^{0}\sigma _{12} \\
\dot{\sigma}_{31} &=&-\Gamma _{31}\sigma _{31}-i\Omega _{p}^{0}e^{-i\phi
}\sigma _{32}-i\Omega _{2}^{0}\sigma _{32^{^{\prime }}}+i\Omega
_{c}^{0}e^{-i\phi }\sigma _{21} \\
&&+i\Omega _{1}^{0}\sigma _{2^{^{\prime }}1} \\
\dot{\sigma}_{22^{^{\prime }}} &=&-\Gamma _{22^{^{\prime }}}\sigma
_{22^{^{\prime }}}+i\Omega _{c}^{0}\sigma _{32^{^{\prime }}}-i\Omega
_{2}^{0}e^{-i\phi }\sigma _{21}-i\Omega _{1}^{0}\sigma _{23} \\
&&+i\Omega _{p}^{0}e^{-i\phi }\sigma _{12^{^{\prime }}} \\
\dot{\sigma}_{21} &=&-\Gamma _{21}\sigma _{21}-i\Omega _{p}^{0}(\sigma
_{22}-\sigma _{11})+i\Omega _{c}^{0}e^{i\phi }\sigma _{31} \\
&&-i\Omega _{2}^{0}e^{i\phi }\sigma _{22^{^{\prime }}} \\
\dot{\sigma}_{2^{^{\prime }}1} &=&-\Gamma _{2^{^{\prime }}1}\sigma
_{2^{\prime }1}-i\Omega _{2}^{0}(\sigma _{2^{^{\prime }}2^{^{\prime
}}}-\sigma _{11})+i\Omega _{1}^{0}\sigma _{31} \\
&&-i\Omega _{p}^{0}e^{-i\phi }\sigma _{2^{^{\prime }}2}
\end{eqnarray*}
where we have used the closed relation of the population $1=\rho _{11}+\rho
_{22}+\rho _{2^{^{\prime }}2^{^{\prime }}}+\rho _{33}$, and $\rho _{jk}=\rho
_{kj}^{*},k\neq j$ to obtain other elements with unlisted equations. We also
defined $\gamma _{2^{^{\prime }}3}=\gamma _{1},\gamma _{23}=\gamma
_{c},\gamma _{12^{^{\prime }}}=\gamma _{2},\gamma _{12}=\gamma _{p}$, and
the related effective decay rates are $\Gamma _{32}=\frac{1}{2}(\gamma
_{1}+\gamma _{c}+\gamma _{p})+i\Delta _{c}$, $\Gamma _{32^{^{\prime }}}=%
\frac{1}{2}(\gamma _{1}+\gamma _{c}+\gamma _{2})+i\Delta _{1}$, $\Gamma
_{31}=\frac{1}{2}(\gamma _{1}+\gamma _{c})+i(\Delta _{p}+\Delta _{c})$, $%
\Gamma _{22^{^{\prime }}}=\frac{1}{2}(\gamma _{2}+\gamma _{p})+i(\Delta
_{p}-\Delta _{2})$ and $\Gamma _{21}=\frac{1}{2}\gamma _{p}+i\Delta
_{p},\Gamma _{2^{^{\prime }}1}=\frac{1}{2}\gamma _{2}+i\Delta _{2}$. $\phi
=\phi _{1}+\phi _{2}-\phi _{c}-\phi _{p}$ is the relative phase of the
applied fields. From the definition above the phase $\phi $ can also be
understood as the relative phase between the two two-photon transitions $%
|1\rangle \leftrightarrow |2\rangle \leftrightarrow |3\rangle $ and $%
|1\rangle \leftrightarrow |2^{^{\prime }}\rangle \leftrightarrow |3\rangle $%
. As will shown below the relative phase plays a crucial role in the gain
spectra.

\begin{figure}[tbph]
\centering
\includegraphics[width=11 cm]{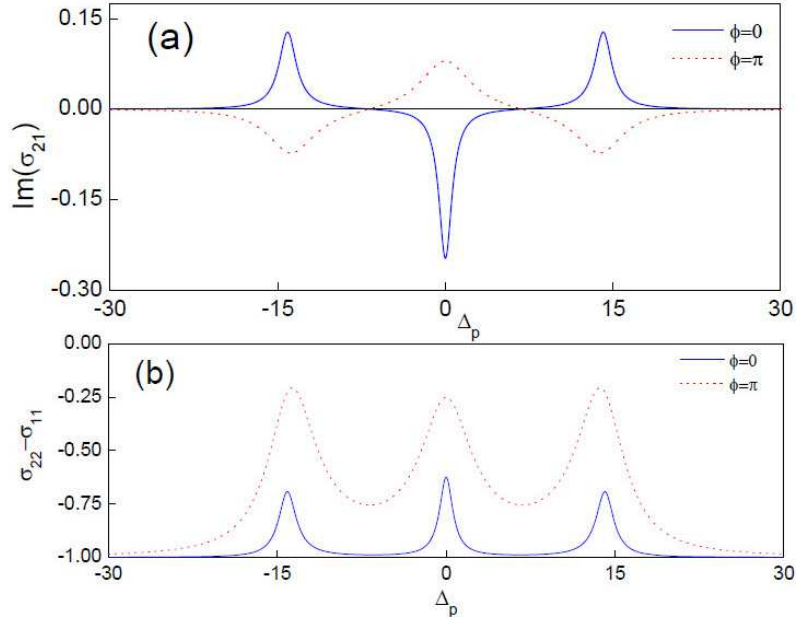}
\caption{(a) Variation of Im($\sigma _{21}$) with probe detuning $\Delta
_{p} $ for $\phi =0$ (solid line) and $\phi =\pi $ (dotted line) for $\Omega
_{c}^{0}$ $=10$, $\Omega _{p}^{0}$ $=0.05$, $\Delta _{c}=\Delta _{1}=0$, $%
\Omega _{1}^{0}=10$ and $\Omega _{2}^{0}=1$. (c) Population difference $%
\sigma _{22}-\sigma _{11}$ vesus probe detuning $\Delta _{p}$ corresponding
to gain in (a).}
\label{fig2}
\end{figure}

Steady state solution of the master equation can be obtained by setting $%
\dot{\sigma}_{ij}=0$. Absorption behavior of the weak probe field $\Omega
_{p}$ is described by Im($\sigma _{21}$). When Im($\sigma _{21}$)$<0$, it
symbols gain behavior. In our calculation we set $\gamma _{j}=\gamma =1$ ($%
j=p,c,1,2$), and scale all the $\Omega _{j}^{0}$ and detunings $\Delta _{j}$
($j=p,c,1,2$) in units of $\gamma $. We choose the probe field to be weak
and real. The main results are shown in Fig. 2.

Phase-dependent gain and corresponding population difference $\sigma
_{22}-\sigma _{11}$ vesus probe detuning $\Delta _{p}$ is shown in Fig. 2.
Im($\sigma _{21}$) vesus probe detuning $\Delta _{p}$ for $\phi =0$ (solid
line) and $\phi =\pi $ (dotted line) is plotted in Fig. 2(a). Other
parameters are chosen as $\Omega _{c}^{0}$ $=10$, $\Omega _{p}^{0}=0.05$, $%
\Delta _{c}=\Delta _{1}=0$, $\Omega _{1}^{0}=10$ and $\Omega _{2}^{0}=1$.
Corresponding population difference $\sigma _{22}-\sigma _{11}$ vesus probe
detuning $\Delta _{p}$ in (a) is shown in Fig. 2(b). Seen from Fig. 2(a), it
is clear that when the relative phase $\phi $ changes from $0$ to $\pi $,
the probe field experiences different gain shapes. Gain and absorption zones
exchange. The most remarkable change lies in the fact that the gain zones
has increased from a single zone near resonant point $\Delta _{p}=0$ to two
separate frequency zones localling around $\Delta _{p}=\pm 15.9$. This means
that the number of gain zones are controlled by the relative phase of the
applied fields. And seen fro Fig. 2(b), population difference $\sigma
_{22}-\sigma _{11}<0$ always holds, which suggests that population inversion
is impossible during the interaction. So the gain behavior does not come
from population inversion between levels $|2\rangle $ and $|1\rangle $.
\begin{figure}[tbph]
\centering
\includegraphics[width=12 cm]{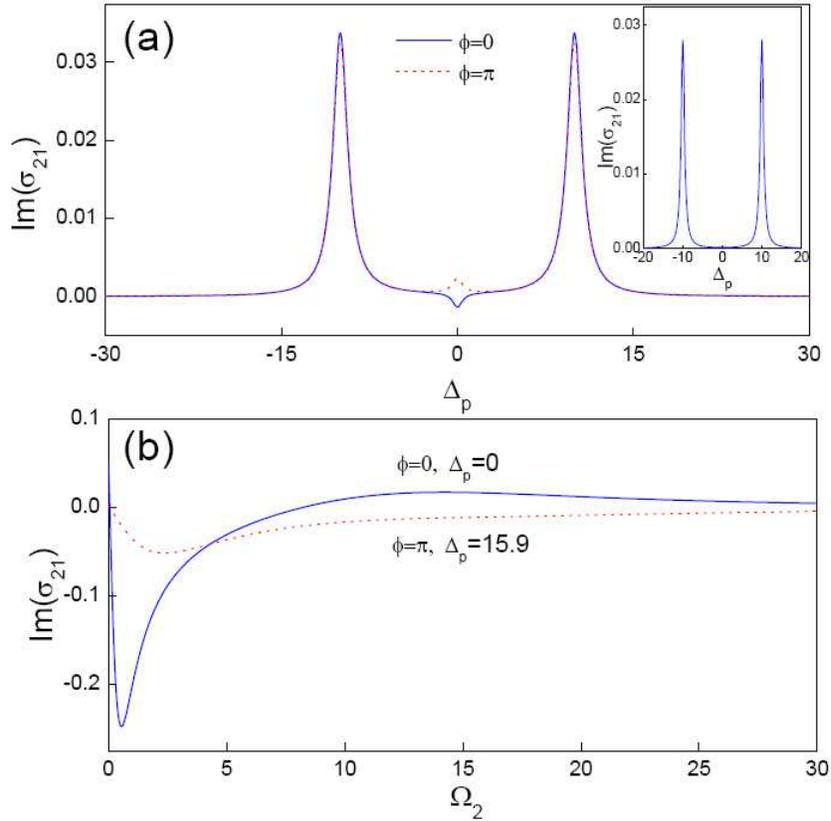}
\caption{(a) Variation of Im($\sigma _{21}$) with probe detuning $\Delta
_{p} $ for $\phi =0$ (solid line) and $\phi =\pi $ (dotted line) for $\Omega
_{c}^{0}$ $=10$, $\Omega _{p}^{0}$ $=0.05$, $\Delta _{c}=\Delta _{1}=0$, $%
\Omega _{1}^{0}=0.1$ and $\Omega _{2}^{0}=0.1$. The inner graph in Fig. 3(a)
is the absorption spectrum vesus $\Delta _{p}$ for three-level cascade
driving without the transitions coupled by the fields $\Omega _{1}$ and $%
\Omega _{2}$. (b) Probe gain versus $\Omega _{2}^{0}$ for $\phi =0,\Delta
_{p}=0$ (solid line) and $\phi =\pi $, $\Delta _{p}=15.9$ (dotted line). The
other parameters are the same as those in Fig. 2(a).}
\label{fig3}
\end{figure}
\begin{figure}[tbph]
\centering
\includegraphics[width=12 cm]{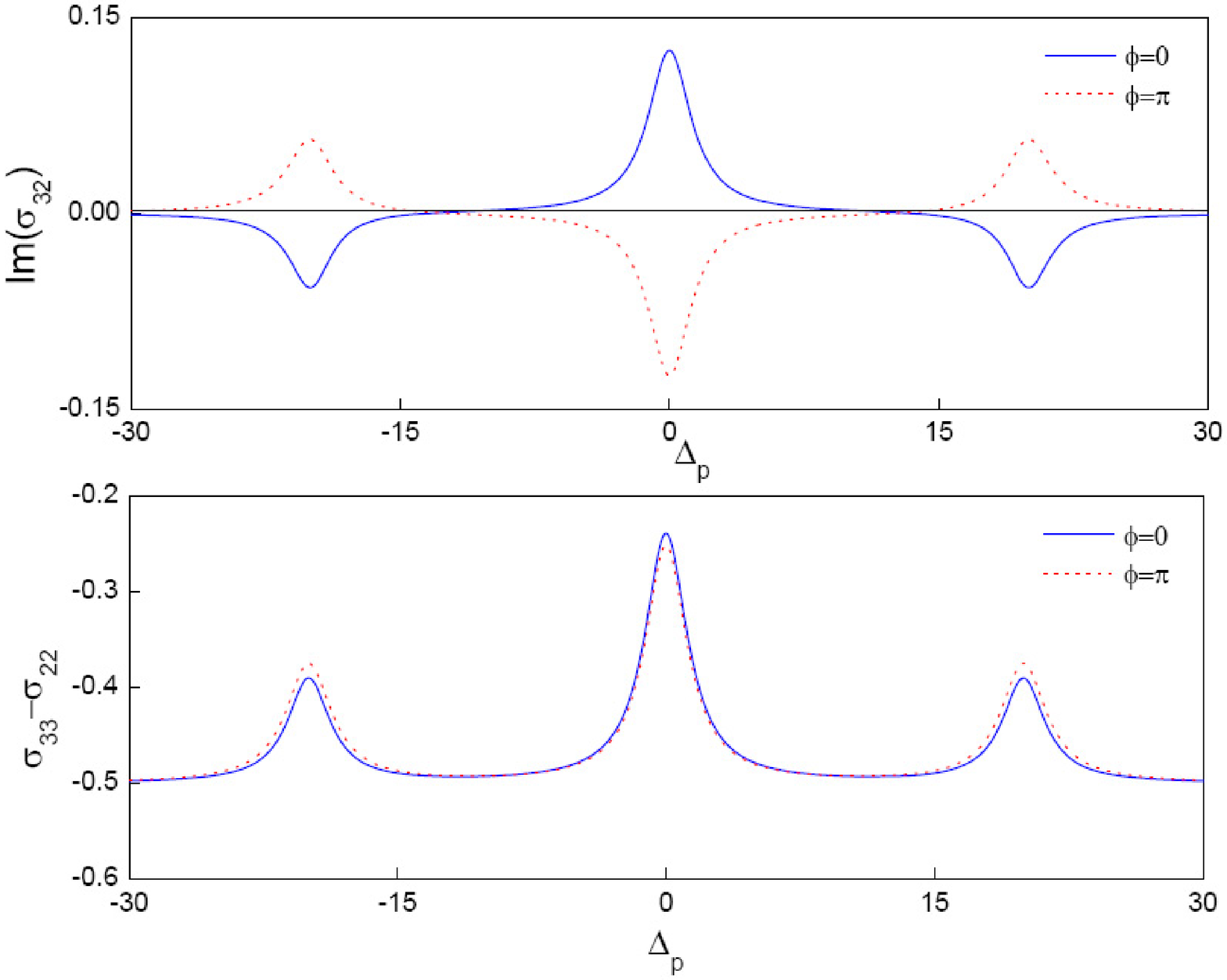}
\caption{(a) Variation of Im($\sigma _{32}$) with probe detuning $\Delta
_{p} $ for $\phi =0$ (solid line) and $\phi =\pi $ (dotted line) for $\Omega
_{c}^{0}$ $=10$, $\Omega _{p}^{0}$ $=0.05$, $\Delta _{c}=\Delta _{1}=0$, $%
\Omega _{1}^{0}=10$ and $\Omega _{2}^{0}=1$ (b) Population difference $%
\sigma _{33}-\sigma _{22}$ vesus probe detuning $\Delta _{p}$ corresponding
to gain in (a).}
\label{fig4}
\end{figure}
Fig. 3(a) shows the case with small driving of one two-photon transition,
i.e., $\Omega _{1}^{0}$ $=\Omega _{2}^{0}$ $=0.1$ for $\phi =0$ (solid
lines) and $\phi =\pi $ (dotted lines) , and the other parameters are the
same as those in Fig. 2(a). Two feathers are presented. One is that the
probe absorption domains with two remarkable absorption peaks. The other is
that when the phase changes from $0$ to $\pi $, the probe gain around the
resonant point goes into the probe absorption while the dominated absorption
keeps unchanged. In order to see the effects of the two two-photon
transitions, we plot the probe absorption for the case of one two-photon
transition $|1\rangle \leftrightarrow |2\rangle \leftrightarrow |3\rangle $
with the same parameters. The inner graph in Fig. 3(a) is the absorption
spectrum vesus $\Delta _{p}$ for three-level cascade driving without the
transitions coupled by the fields $\Omega _{1}$ and $\Omega _{2}$. The
results show that electromagnetically induced transparency can be obtained
and no phase dependence and no gain are found for single two-photon
transition. Thus it is the two two-photon transitions is the origin of the
probe gain and the phase dependence. In Fig. 3(b) the probe absorption Im($%
\sigma _{21}$) vesus $\Omega _{2}^{0}$, the amplitude of $\Omega _{2}$ is
also plotted for $\phi =0,\Delta _{p}=0$ (solid line) and $\phi =\pi $, $%
\Delta _{p}=15.9$ (dotted line). The other parameters are the same as those
in Fig. 2(a). It is easy to see that the parameters we choose in Fig. 2(a)
are optimal for gain behavior.
\begin{figure}[tbph]
\centering
\includegraphics[width=12 cm]{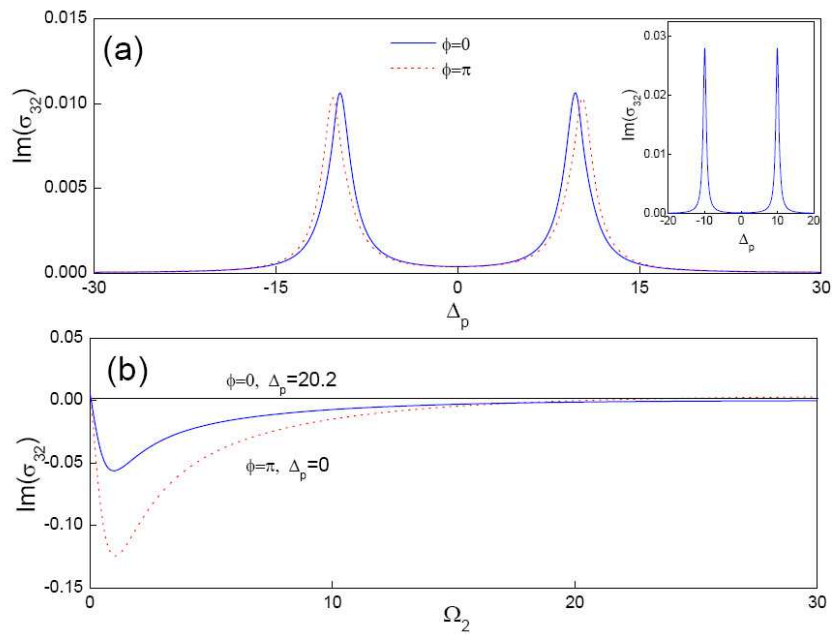}
\caption{(a) Variation of Im($\sigma _{32}$) with probe detuning $\Delta
_{p} $ for $\phi =0$ (solid line) and $\phi =\pi $ (dotted line) for $\Omega
_{c}^{0}$ $=10$, $\Omega _{p}^{0}$ $=0.05$, $\Delta _{c}=\Delta _{1}=0$, $%
\Omega _{1}^{0}=0.1$ and $\Omega _{2}^{0}=0.1$. The inner graph in Fig. 5(a)
is the absorption spectrum vesus $\Delta _{p}$ for three-level cascade
driving without the transitions coupled by the fields $\Omega _{1}$ and $%
\Omega _{2}$. (b) Probe gain versus $\Omega _{2}^{0}$ for $\phi =0,\Delta
_{p}=20.2$ (solid line) and $\phi =\pi $, $\Delta _{p}=0$ (dotted line). The
other parameters are the same as those in Fig. 4(a).}
\label{fig5}
\end{figure}
When we exchange the probe field $\Omega _{p}$ and the driving field $\Omega
_{c}$ with coupling scheme shown in Fig. $1$(b), gain without inversion can
also be obtained. We call this case up-level probing. The results are
presented in Fig. 4 and Fig. 5 with similar parameters as those in Fig. 2
and Fig. 3. It is clear that gain without inversion also exhibits in such a
system and the phase dependence of the gain on relative phase $\phi $ is
presented as well. There are two differences between the low level probing
and the up level probing case. One is that the number of gain zone is
different with the same phase. For specific, when $\phi =0$, the low-level
coupling has single gain without inversion zone at $\Delta _{p}=0$ while the
up-level coupling has two gain without inversion zones at $\Delta _{p}=\pm
20.2$; when $\phi =\pi $, the low-level coupling has two gain without
inversion zones at $\Delta _{p}=\pm 15.9$ while the up-level coupling has
only one gain without inversion zones at $\Delta _{p}=0$. The other feature
is that no gain is presented when the two-photon transition $|1\rangle
\leftrightarrow |2^{^{\prime }}\rangle \leftrightarrow |3\rangle $ is small
when the phase changes from $0$ to $\pi $ and the phase just makes the shift
of the absorption a little. In a word the absorption spectra act like the
probe absorption for the case of one two-photon transition $|1\rangle
\leftrightarrow |2\rangle \leftrightarrow |3\rangle $ with the same
parameters except the phase influence. In Fig. 5(b) the dependence of probe
gain on $\Omega _{2}^{0}$ is also presented. From this, one can see that the
optimal parameters are used in Fig. 4(a).

To under stand the above results, we can simply use the steady state
expression of two terms as following
\begin{eqnarray}
\mathop{\rm Im}%
(\sigma _{21}) &=&\frac{1}{D}%
\mathop{\rm Im}%
[(\frac{\gamma _{p}}{2}-i\Delta _{p})(i\Omega _{c}^{0}\sigma _{31}-i\Omega
_{2}^{0}\sigma _{22^{^{\prime }}})e^{i\phi }]  \nonumber \\
&&-\frac{\frac{1}{2}\Omega _{p}^{0}\gamma _{p}(\sigma _{22}-\sigma _{11})}{%
\frac{1}{4}\gamma _{p}^{2}+\Delta _{p}^{2}} \\
\mathop{\rm Im}%
(\sigma _{32}) &=&\frac{1}{D}%
\mathop{\rm Im}%
[(\frac{\gamma _{c}}{2}-i\Delta _{c})(-i\Omega _{p}^{0}e^{-i\phi }\sigma
_{31}+i\Omega _{1}^{0}\sigma _{2^{^{\prime }}2})]  \nonumber \\
&&-\frac{\frac{1}{2}\Omega _{c}^{0}\gamma _{c}(\sigma _{33}-\sigma _{22})}{%
\frac{1}{4}\gamma _{c}^{2}+\Delta _{c}^{2}}
\end{eqnarray}
where $D=\frac{1}{4}\gamma _{p}^{2}+\Delta _{p}^{2}$ . From eq. (6), it is
easy to see that when no population inversion happens, i.e., $(\sigma
_{22}-\sigma _{11})<0$, the first term is always positive. So when gain
occurs (Im$(\sigma _{21})<0$), the second term contributes. Similar results
also hold for up-level probing case just by make the replace of $\sigma _{21}
$ by $\sigma _{32}$ and exchange the field $\Omega _{c}$ and $\Omega _{p}$.
Due to pure absorption and no phase-dependent behavior of three-level
cascade driving $|1\rangle \leftrightarrow |2\rangle \leftrightarrow
|3\rangle $ by $\Omega _{c}$ and $\Omega _{p}$, we can conclude that it is
the two-photon transition $|1\rangle \leftrightarrow |2^{^{\prime }}\rangle
\leftrightarrow |3\rangle $ driving by $\Omega _{1}$ and $\Omega _{2}$ that
induces the gain without inversion and the two two-photon transitions $%
|1\rangle \leftrightarrow |2(2^{^{\prime }})\rangle \leftrightarrow
|3\rangle $ loop structure induces the phase dependence of the gain behavior.

In conclusion, gain without inversion in a four level loop-transition atomic
system has been investigated. The main results are two features: (i) Gain
without inversion is exhibited by varing the amplitude of coupled fields.
(ii) Gain without inversion displays a phase dependence on the relative
phase between the fields applied on the two two-photon transitions. As the
phase changes from 0 to $\pi $, gain and absorption zones exchange.
Population inversionless holds for all the case. Gain without inversion and
phase-dependence are attributed to interference induced by such a
loop-transition structure.

{\bf Acknowledgments}
This work is supported by the Scientific Research Plan of the Provincial Education Department in Hubei (Grant No. Q20101304) and NSFC under Grant No. 11147153.

\end{document}